\documentclass[aps,pra,twocolumn,showpacs,superscriptaddress,letterpaper]{revtex4-1}

\usepackage{graphicx}
\usepackage{mdframed}
\usepackage{framed}
\usepackage{indentfirst}
\usepackage{natbib}
\usepackage{amsmath,amssymb}
\usepackage{subfigure}
\usepackage{enumitem}

\begin{document}

\title{Correlating students' beliefs about experimental physics with lab course success}


\author{Bethany R. Wilcox}
\affiliation{Department of Physics, University of Colorado, 390 UCB, Boulder, CO 80309}

\author{H. J. Lewandowski}
\affiliation{Department of Physics, University of Colorado, 390 UCB, Boulder, CO 80309}
\affiliation{JILA, University of Colorado Boulder, Boulder, CO 80309}

\begin{abstract}
Student learning in instructional physics labs is a growing area of research that includes studies exploring students' beliefs and expectations about experimental physics.  To directly probe students' epistemologies about experimental physics and support broader lab transformation efforts both at the University of Colorado Boulder (CU) and nationally, we developed the Colorado Learning Attitudes about Science Survey for Experimental Physics (E-CLASS).  Previous work focused on establishing the accuracy and clarity of the instrument through student interviews and preliminary testing.  Ongoing validation efforts include establishing the extent to which student epistemologies as measured by E-CLASS align with other measures of student learning outcomes (e.g., course grades).  Here, we report on correlations between final course grades and E-CLASS scores from two semesters of introductory and upper-division lab courses at CU and discuss implications of our findings for the validity of the E-CLASS instrument.
\end{abstract}

\maketitle

\section{\label{sec:intro}Introduction}

The Physics Education Research (PER) community has a growing body of research dedicated to investigating students' attitudes and epistemologies about what it means to know, learn, and do physics.  In addition to having implications for retention and persistence, a student's epistemological stance can potentially impact their content learning in a physics course \cite{lising2005epistemology}.  Several assessments have been developed to measure students' epistemologies and expectations (E\&E) both about physics specifically \cite{adams2006class,redish1998mpex} and the nature of science more generally \cite{halloun1996vass, abd2001vnos, chen2006vose, elby2001ebaps}.  

In developing standardized assessment instruments, there are a number of distinct aspects of validity and reliability that must be established \cite{engelhardt2009ctt}.  For example, convergent validity is determined by comparing the scores on the assessment with other, related student outcomes \cite{adams2011development}.  For conceptual assessments, convergent validity is typically established relative to students' final course grades.  However, for E\&E assessments, it is reasonable to ask if we expect the same level of correlation with course performance \cite{redish1998mpex,hammer1989two}, particularly given that promoting expert-like attitudes and beliefs is rarely an \emph{explicit} goal of physics courses.  Of the available E\&E assessments, only the VASS (Views About Science Survey) has published results reporting a modest but significant correlation ($r\sim0.3$) between VASS score and final course grades in high school and college physics courses \cite{halloun2001vass}.  Weak, but statistically significant, correlations have also been reported between students' scores on either the CLASS (Colorado Learning Attitudes about Science Survey) or MPEX (Maryland Physics Expectations Survey) and their gains on various validated conceptual assessments \cite{perkins2005class, pollock2005reforms, kortemeyer2007mpex}.  Literature on other assessments has focused instead on demonstrating that populations who can be expected to have more expert-like epistemologies (e.g., faculty and graduate students) score higher on E\&E instruments than novices \cite{redish1998mpex}.  

The goal of the work described here is to further explore the concept of convergent validity as it pertains to E\&E surveys by examining the relationship between students' overall course grades and their scores on one specific instrument -- the Colorado Learning Attitudes about Science Survey for Experimental Physics (E-CLASS) \cite{zwickl2014eclass}.  E-CLASS is a 30 item, Likert-style survey designed to measure students' epistemologies and expectations about experimental physics.  The development and initial validation of the E-CLASS is described in Ref.\ \cite{zwickl2014eclass}.  E-CLASS was developed to support upper-division laboratory course transformation efforts at the University of Colorado Boulder (CU) \cite{zwickl2013adlab} as well as similar efforts nation-wide.  This work is part of ongoing research into the overall validity and reliability of the E-CLASS.

\section{\label{sec:methods}Methods}

Data for this study were collected from two semesters of the four core physics laboratory courses at CU.  These courses are described in Table \ref{tab:courses} and span both the lower- and upper-division level.  E-CLASS was administered online to each of the courses both at the beginning and end of each semester as a normal part of the class.  An example item from the survey is given in Fig.\ \ref{fig:exampleQ}.  Students were awarded only participation credit for completing the survey.  Final letter grades were collected for all students who agreed to release their grade data, and students were assigned a standard grade point value for each letter grade (i.e., $A=4.0$, $A-=3.7$, $B+=3.3$, $B=3.0$, etc.).  Only students with matched post-test ECLASS scores and final course grades were included in the following analysis.  The E-CLASS also includes a filtering question that prompts the students to select `agree,' not `strongly agree,' in order to eliminate responses from students who did not read the questions.  Students in the final, matched data set had, on average, final course scores that were 0.22 grade points higher than the course overall.  We anticipated this type of selection effect because survey completion was worth only minimal credit for the students; however, the difference in final grade between the participants and the class overall is small and thus we take our sample to be reasonably representative.  

\begin{figure}
\begin{minipage}{\linewidth}
\begin{mdframed}
\vspace{1mm}When doing an experiment, I try to understand how the experimental setup works. \vspace{1mm}\\
\emph{What do YOU think when doing experiments for class?}
\begin{center}Strongly disagree 1 \hspace{1mm}2 \hspace{1mm}3 \hspace{1mm}4 \hspace{1mm}5 Strongly agree\end{center}
\emph{What would experimental physicists say about their research?}
\begin{center}Strongly disagree 1 \hspace{1mm}2 \hspace{1mm}3 \hspace{1mm}4 \hspace{1mm}5 Strongly agree\end{center}
\end{mdframed}
\end{minipage}
\caption{An example item from the E-CLASS.  Students are asked to indicate their level of agreement with the provided statement from their own perspective and that of an experimental physicist.   }\label{fig:exampleQ}
\end{figure}

While the survey was administered both pre and post in all courses, the following analysis focuses exclusively on post-test scores rather than pre-post shifts, as our primary goal was to validate students' scores on the instrument rather than to compare the impact of different courses or interventions.  We will also limit our analysis to students' personal beliefs, rather than their prediction of what an experimental physicist would say (see Fig.\ \ref{fig:exampleQ}).  The overall post-test score for each student is given by the fraction of items that they answered favorably (i.e., consistent with established expert responses).  For analysis of individual items rather than overall scores, students' responses to each 5-point Likert item are condensed into a standardized, 3-point scale in which the responses `(dis)agree' and `strongly (dis)agree' are collapsed into a single category.  Thus, student responses to individual items are coded simply as favorable, neutral, or unfavorable.  The collapsing of the 5-point scale to 3-points is common in analysis of Likert-style items and is motivated by the inherently ordinal, rather than interval, nature of the Likert response scale \cite{lovelace2013attitudes}.  Here, we further collapse neutral and unfavorable responses to a single category so that, consistent with the overall E-CLASS score, student responses to each item are classified simply as favorable (+1) or not-favorable (0).

\section{\label{sec:results}Results}

Aggregating across all students in all courses (N=873), we found an overall correlation coefficient of $r=0.05$ between final course grade and E-CLASS score (fraction of items answered favorably).  This correlation is neither practically or statistically significant ($p=0.1$).  However, it is also reasonable to expect that this correlation might vary between courses.  In particular, the first year lab at CU is a large service course catering primarily to engineering, rather than physics majors.  Thus, the learning goals and grading practices of this course are not necessarily aligned with the learning goals targeted by E-CLASS, which were developed in collaboration with physics faculty to capture their desired learning outcomes for physics students in their upper-division lab courses.  

\begin{table}
\caption{Core physics lab courses offered at CU.  Valid N represents the number of students for which we have matched post E-CLASS scores and final grades.  All courses except $^\dagger$ were offered twice in the two semesters of data collection. }\label{tab:courses}
\begin{tabular}{l l c c}
\hline
\hline
Course & Year\hspace{1mm} & N & Response\\
 & & & Rate \\
\hline
Experimental Physics 1 & 1st & 717 & 65\% \\
Experimental Modern Physics& 2nd & 76 & 52\% \\
Electronics for Physical Sciences \hspace{2mm} & 3rd & 64 & 82\% \\
Advanced Laboratory $^\dagger$& 4th & 16 & 100\% \\
\hline
\hline
\end{tabular}
\end{table}

To investigate this potential variability across courses, we divided the students into two subgroups composed of those in the first-year lab (N=717) and those in the second-, third-, and fourth-year labs (N=156), which we will refer to as the beyond-first-year (BFY) labs.  Our motivation for dividing the students in this way was threefold.  Firstly, it provides a clear distinction between the classes that is applicable beyond CU.  Secondly, it preserves sufficient statistical power given the significantly smaller size of the BFY courses. Finally, there is a significant shift in the student population between the first year and BFY labs.  The first year lab is taken by most STEM majors including all engineering majors, whereas the BFY courses are taken almost exclusively by physics, engineering physics, and astrophysics majors.  

For students in the first-year lab, the correlation between overall E-CLASS score and final grade is small and not statistically significant ($r=0.01$, $p=0.7$).  However, for the BFY labs, this correlation increases to $r=0.19$ and is statistically significant ($p=0.02$).  This correlation, while still weak, is similar in magnitude to the correlations reported between CLASS/MPEX scores and conceptual learning gains as measured by conceptual assessments such as the Force Concept Inventory \cite{perkins2005class, pollock2005reforms, kortemeyer2007mpex}.  We are not arguing that the relationship between E-CLASS scores and final grades is a causal one; however, these results do suggest that the link between course performance and epistemological stance is stronger in more advanced lab courses than in the first-year lab.  

\begin{figure*}
\includegraphics{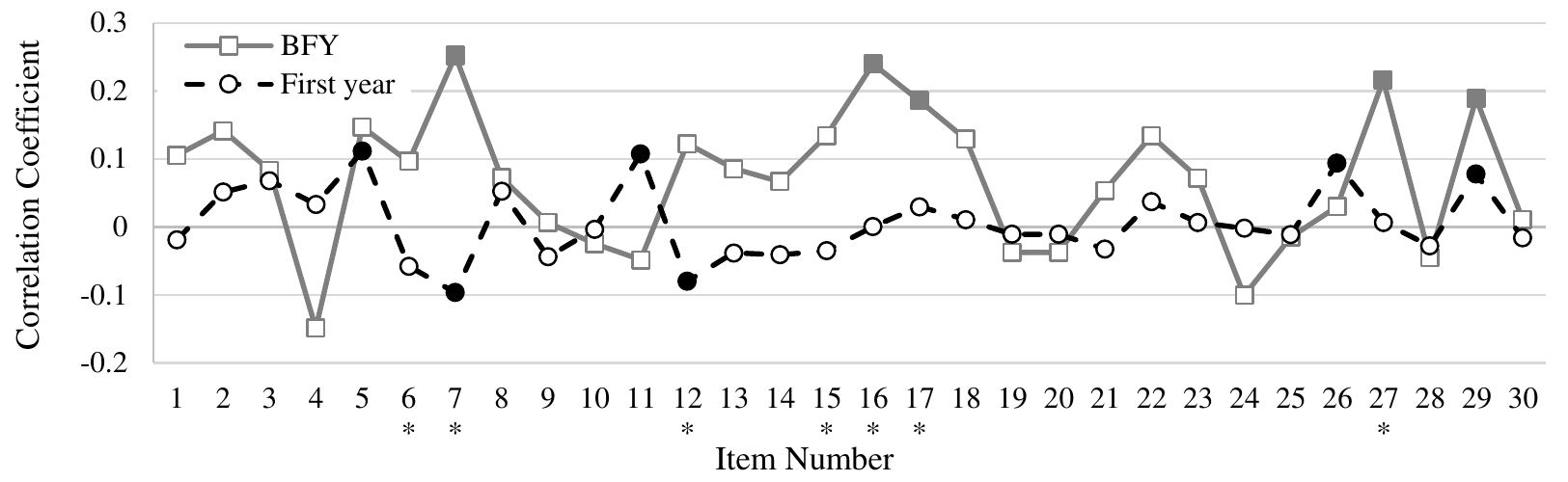}
\caption{Item score to final grade correlations for each of the 30 E-CLASS items.  Statistically significant correlations ($p<0.05$) are indicated by solid markers.  Items for which there is a statistically significant difference ($p<0.05$) between the correlations for the first-year and the BFY courses are indicated with an asterisk.  For a list of individual item prompts see Ref.\ \cite{zwickl2014eclass}. }\label{fig:itemcorrel}
\end{figure*}

We can also break down student responses to E-CLASS for each of the 30 items individually to examine which of these items best correlates with grades, and investigate how this varies by course level.  For example, one item on the E-CLASS asks students to agree or disagree with the statement ``Scientific journal articles are helpful for answering my own questions and designing experiments.''  Most instructors would neither expect their first-year lab students to agree with this statement nor consider this to be one of the goals of their courses.  Thus, we would reasonably expect variability in item correlations across both items and courses.  A plot of the item-grade correlations for each item on the E-CLASS is given in Fig.\ \ref{fig:itemcorrel}.  Fig.\ \ref{fig:itemcorrel} also indicates both items for which there is a statistically significant item-grade correlation and items for which there is a statistically significant difference between the correlations for first-year and BFY courses.  

\begin{figure*}
\includegraphics[width=6in]{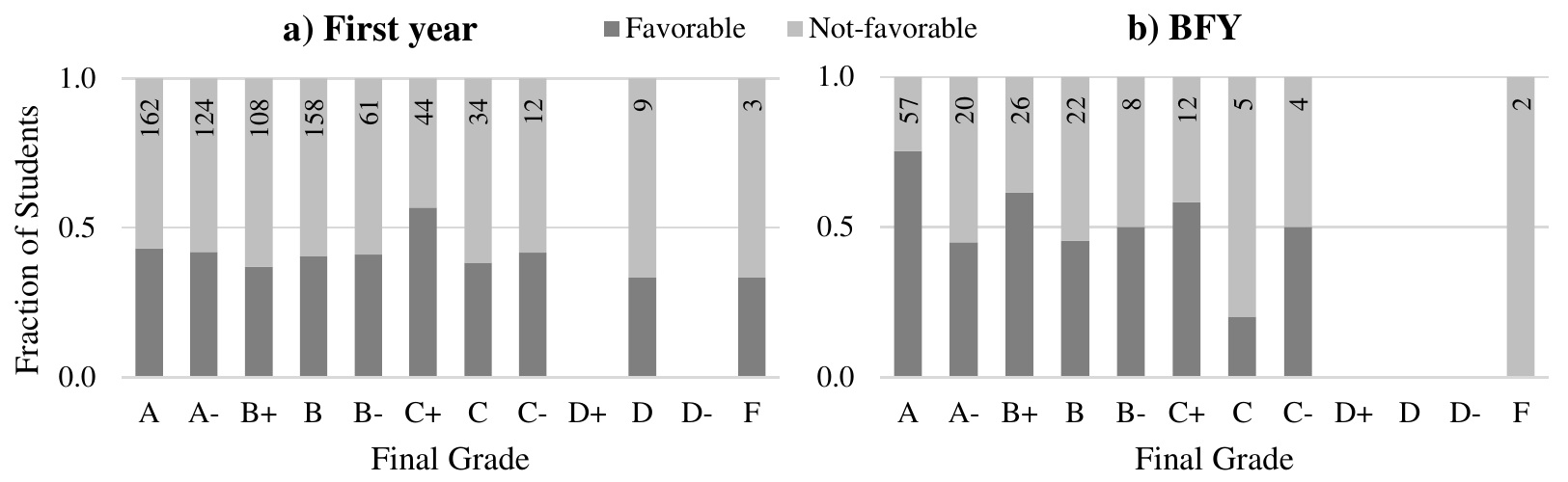}
\caption{Fraction of students who answered favorably vs. not-favorably to Q16 for each letter grade in: a) the first-year lab and b) the beyond-first-year labs.  Number of students who received each final letter grade is given at the top of each bar.  }\label{fig:Q16dist}
\end{figure*}

There are multiple reasons why an item might show high or low correlation with overall grade; thus, on its own, the correlation coefficient between item score and final grade (Fig.\ \ref{fig:itemcorrel}) provides limited insight into the nature of the relationship between these two measures.  To understand these correlations and the differences between first-year and BFY courses, we must examine the distribution of students' final grades relative to their responses to each item.  As an example of this, we focus on Q16, which asks students to rate their agreement with the statement ``\emph{The primary purpose of doing physics experiments is to confirm previously known results}.''  Fig.\ \ref{fig:Q16dist} provides the fraction of students who answered favorably and not-favorably for each final letter grade in both the first-year and BFY courses.   

From Fig.\ \ref{fig:Q16dist} we can see that a significant fraction of the students in both the first-year and BFY labs do not respond favorably to this question.  However, for the first-year lab, the fraction of favorable responses only fluctuates by roughly 10\% based on course performance, whereas there is a general downward trend to the fraction of favorable responses as final grade decreases for the BFY labs.  This result is consistent with the types of experiments done in these courses.  The first-year lab at CU involves almost exclusively confirmatory labs in which the students measure the value of a previously known parameter.  Alternatively, the labs in the BFY courses are more exploratory and design-based.  These courses often also include a final project that is selected and designed by the students.  

Another question that stands out as a potentially interesting case study is Q7, which asks students to rate their agreement with the statement ``\emph{I don't enjoy doing physics experiments}.''  A plot like that in Fig.\ \ref{fig:Q16dist} shows that in the first-year lab, high performing students (final grades between A and B-) were evenly split between favorable and not-favorable responses, while the lower performing students (final grades between C+ and F) had a slightly higher fraction of favorable responses. This result may be partially explained by the student population of this course.  A large fraction of the students were STEM (not physics) majors fulfilling a required natural science credit.  These students are often motivated not by an enthusiasm for physics, but rather the desire to get a good grade and continue onto their major courses or complete their degree.  Alternatively, in the BFY labs, the fraction of students with favorable responses was highest for the A students and had a general downward trend as final grades decreased.   As students in these courses were primarily in physics or physics-related majors, they are more likely to be intrinsically motivated by an enthusiasm for physics and/or discouraged by poor performance.

\section{\label{sec:discussion}Discussion}

We examined the correlation between students' overall E-CLASS score and their final grade for two semesters of four lab courses at CU.  We found that while there was no statistically significant correlation between grades and E-CLASS score for the student overall, there was a small but significant correlation of $r=0.19$ for the students in the beyond-first-year labs.  However, $r=0.19$ still represents a fairly weak correlations, suggesting that E-CLASS scores are not good predictors of students' course performance (or vice versa).  To interpret these findings, we need to take a closer look at the way final grades are determined for these courses.  In the first-year lab course at CU, 60\% of students' grades are based on their lab reports, while the remainder is a combination of participation, traditional homework, and pre-lab assignments.  Lab reports are graded on clarity of both presentation and physics content, including data presentation, data analysis, and discussion of uncertainty.  

Grading schemes in the BFY courses, while varied, are generally characterized by roughly 50-80\% of a students' final grade  stemming from performance on lab notebooks, lab reports, and, for the 3rd and 4th year courses, the final project.  The final project is typically a multi-week group project in which the students propose their own project topics rather than being assigned one or choosing from a predetermined list.  The open-ended nature of these projects make them one place where we might expect to see clearer connections between students' epistemologies and their successes and failures when completing their project.  Indeed, we might hypothesize that the inclusion of the final project may account for the marginal increase in the correlation between E-CLASS score and final grade for the BFY courses.  However, our results still suggest that students' epistemological knowledge and beliefs are not being effectively captured by their final course score.  If we, as laboratory instructors, value the epistemological development of our students, we need to begin measuring that development explicitly through our grading practices.

The validation of the E-CLASS is ongoing.  Current work includes utilizing the growing, national data set of student responses from multiple courses and levels across many institutions to establish the statistical validity and reliability of the instrument for a broad student population.  This national data set may also help us to determine which grading practices best align with the goal of promoting students' epistemological development.  Future work could include longitudinal studies of the change in students epistemologies as they advance through the physics curriculum, and further investigation of the link between students' epistemology and their success on project-based assignments by correlating student's E-CLASS scores with their scores on the final project in the BFY labs.  
\vspace{-2pt}

\begin{acknowledgments}
This work was funded by the NSF-IUSE Grant DUE-1432204.  Special thank you to Dimitri Dounas-Frazer and PER@C for all their help and feedback.  
\end{acknowledgments}

\bibliography{master-refs-ECLASS-6-15-v2-mod}

\end{document}